\documentclass[conference,9pt]{IEEEtran}
\IEEEoverridecommandlockouts
\usepackage{cite}
\usepackage{amsmath,amssymb,amsfonts}
\usepackage{algorithm}
\usepackage{graphicx}
\usepackage{textcomp}
\usepackage{xcolor}

\usepackage{booktabs}
\usepackage[hidelinks]{hyperref}
\usepackage{siunitx}
\usepackage{mathtools}
\usepackage{comment}
\usepackage{multirow}
\usepackage{makecell}

\usepackage{color}
\usepackage{multirow}
\usepackage{algpseudocode}
\usepackage{url}
\usepackage{amssymb,amsmath,bm}
\usepackage{textcomp}
\usepackage{cite}
\usepackage{CJKutf8}
\usepackage{hhline}
\usepackage{bbding}
\usepackage{pifont}
\usepackage{wasysym}
\usepackage{arydshln}
\usepackage{tikz}
\usetikzlibrary{calc}
\usetikzlibrary{tikzmark}
\usetikzlibrary{arrows, positioning,intersections,angles}
\usetikzlibrary{fit}
\usetikzlibrary{patterns}

\makeatletter
\let\MYcaption\@makecaption
\makeatother
\usepackage[font=footnotesize,subrefformat=parens]{subcaption}
\makeatletter
\let\@makecaption\MYcaption
\makeatother

\usepackage{xspace}
\makeatletter
\DeclareRobustCommand\onedot{\futurelet\@let@token\@onedot}
\def\@onedot{\ifx\@let@token.\else.\null\fi\xspace}

\makeatother

\def\equationautorefname~#1\null{(#1\null)}

\renewcommand{\sectionautorefname}{Section}
\renewcommand{\subsectionautorefname}{\sectionautorefname}

\def\L{{\cal L}}
\newcommand{\bhline}[1]{\noalign{\hrule height #1}}

\newcommand{\maru}[1]{\raise0.2ex\hbox{\textcircled{\scriptsize{#1}}}}

\let\orgautoref\autoref

\renewcommand{\autoref}[1]
{%
\def\figureautorefname{Fig.}%
\def\subfigureautorefname{\figureautorefname}%
\def\sectionautorefname{Sec.}%
\def\subsectionautorefname{\sectionautorefname}%
\def\subsectionautorefname{\sectionautorefname}%
\orgautoref{#1}%
}

\def\appendixautorefname~#1\null{~#1 \null}

\makeatletter
\newcommand{\figcaption}[1]{\def\@captype{figure}\caption{#1}}
\newcommand{\tblcaption}[1]{\def\@captype{table}\caption{#1}}
\makeatother

\makeatletter % changes the catcode of @ to 11
\newcommand{\linebreakand}{%
  \end{@IEEEauthorhalign}
  \hfill\mbox{}\par
  \mbox{}\hfill\begin{@IEEEauthorhalign}
}
\makeatother % changes the catcode of @ back to 12

\def\BibTeX{{\rm B\kern-.05em{\sc i\kern-.025em b}\kern-.08em
    T\kern-.1667em\lower.7ex\hbox{E}\kern-.125emX}}
\begin{document}

\title{\fontsize{19.7pt}{24pt}\selectfont Alignment-Free Training for Transducer-based Multi-Talker ASR}

%\title{Alignment-Free Training for Transducer-based Multi-Talker ASR}

%\author{\IEEEauthorblockN{Takafumi Moriya, Shota Horiguchi, Ryo Masumura, Takanori Ashihara, Hiroshi Sato, Kohei Matsuura, Masato Mimura}
\author{\IEEEauthorblockN{Takafumi Moriya, Shota Horiguchi, Marc Delcroix, Ryo Masumura, \\ Takanori Ashihara, Hiroshi Sato, Kohei Matsuura, Masato Mimura}
\IEEEauthorblockA{NTT Corporation, Japan}}

\maketitle

\begin{abstract}
Extending the RNN Transducer (RNNT) to recognize multi-talker speech is essential for wider automatic speech recognition (ASR) applications. Multi-talker RNNT (MT-RNNT) aims to achieve recognition without relying on costly front-end source separation. MT-RNNT is conventionally implemented using architectures with multiple encoders or decoders, or by serializing all speakers' transcriptions into a single output stream. The first approach is computationally expensive, particularly due to the need for multiple encoder processing. In contrast, the second approach involves a complex label generation process, requiring accurate timestamps of all words spoken by all speakers in the mixture, obtained from an external ASR system. In this paper, we propose a novel alignment-free training scheme for the MT-RNNT (MT-RNNT-AFT) that adopts the standard RNNT architecture. The target labels are created by appending a prompt token corresponding to each speaker at the beginning of the transcription, reflecting the order of each speaker's appearance in the mixtures. Thus, MT-RNNT-AFT can be trained without relying on accurate alignments, and it can recognize all speakers' speech with just one round of encoder processing. Experiments show that MT-RNNT-AFT achieves performance comparable to that of the state-of-the-art alternatives, while greatly simplifying the training process. 
\end{abstract}

\begin{IEEEkeywords}
speech recognition, end-to-end, neural transducer, multi-talker, alignment-free training
\end{IEEEkeywords}

\section{Introduction}
RNN Transducer (RNNT)~\cite{Graves2012} is promising for streaming automatic speech recognition (ASR)~\cite{tara2020rnntdevice}, but it struggles to handle multi-speaker overlapped inputs. 
To address this, a variety of multi-talker RNNT (MT-RNNT) methods have been proposed to transcribe the overlapping speech of multiple speakers~\cite{SklyarPL21PIT,Sklyar2022mtMSRNNT,sklyar2022sep-rnnt,lu2021surtorg,Lu2021surt,lu2022epsurt,raj2022mtrnnt,raj2024surt2,raj2024surt_hat,Kanda2022mtasr,kanda2022SAtSOT,muqiao2023tsot_simulated,jian2024tsot_lmadapt}.

Several MT-RNNT approaches employ multiple encoder and/or decoder branches with permutation invariant training (PIT)~\cite{SklyarPL21PIT,Sklyar2022mtMSRNNT,sklyar2022sep-rnnt} or heuristic error assignment training (HEAT)~\cite{lu2021surtorg,Lu2021surt,lu2022epsurt,raj2022mtrnnt,raj2024surt2,raj2024surt_hat}.
Although these MT-RNNTs do not use any front-end speech separation, decoding the speech of all speakers is often computationally intensive. 
This is because the encoder processing, which is the most computationally demanding operation~\cite{saon2020alsd,fasoli2022lmfusiontime,yongqiang2023rnnt_joint_ctc_training}, must be performed individually for each speaker in the mixture. This significantly increases the computational costs in both training and decoding. This issue poses a critical challenge to streaming applications, and thus it is preferable to have a single encoder that can simultaneously recognize multi-talker inputs. 

To achieve this, token-level serialized output training (tSOT) for MT-RNNT (MT-RNNT-tSOT) was proposed~\cite{Kanda2022mtasr}. In tSOT, multiple speakers' transcriptions are serialized into a single output stream based on the order of subword-level occurrence timestamps, regardless of the speakers. This enables multi-talker ASR with the standard RNNT architecture. 
However, the serialization requires accurate timestamps for each token, which must be obtained through forced alignment from an external ASR system.
Moreover, performing forced alignment on real recording mixtures is particularly challenging, and low-quality alignments result in the degradation of MT-RNNT-tSOT performance.

In this paper, we propose a novel MT-RNNT training scheme that retains the standard RNNT architecture while significantly simplifying the training process, without requiring any alignment. 
We refer to this approach as MT-RNNT with \textit{alignment-free training} (MT-RNNT-AFT). 
For MT-RNNT-AFT, we introduce a prompt token that specifies the order of speakers' appearances in the mixture. 
The target labels for each speaker are then created by simply appending the prompt token at the beginning of each transcription. 
The losses are computed individually between the target label and the prediction for each speaker, and then summed. 
MT-RNNT-AFT can decode all speakers' speech in a first-in-first-out manner, requiring just one round of encoder processing. 
The decoder can simultaneously recognize all speakers' speech by batching its processing~\cite{saon2020alsd} for all speakers, which is made possible by the use of identical parameters. 
The computational costs are much lower than MT-RNNT which requires distinct encoder outputs for all speakers, as mentioned above. 

MT-RNNT-AFT can output each speaker's hypothesis individually, unlike MT-RNNT-tSOT, which outputs a single serialized transcription in a more complex format.
Therefore, MT-RNNT-AFT can utilize various effective approaches developed for standard single-talker ASR.
Leveraging this advantage, we introduce self-knowledge distillation (KD) using parallel single/multi-talker ASR data. 
We can naturally use the parallel data because the mixture is created on-the-fly using multiple single-talker voices. 
We distill knowledge from the MT-RNNT-AFT outputs, which are generated from single-talker ASR data, to the outputs of MT-RNNT-AFT itself, produced using multi-talker ASR data, similar to~\cite{zhang19i_interspeech,moriya2023tsrnnkd}. 
We also employ language model (LM) integration~\cite{anjuli2018shallowfusion,meng2021ilme} during decoding. 

Experiments demonstrate that MT-RNNT-AFT achieves comparable performance to MT-RNNT-tSOT in offline mode, even though MT-RNNT-AFT does not use any rich alignments from external ASR systems. 
Moreover, KD and LM integration further improve the recognition performance. 
Our best systems match the recognition performance of state-of-the-art alternatives in both streaming and offline modes, while employing a much simpler training scheme.

%%%%%

%\vspace{-0.15cm}
\section{Related work}
%\vspace{-0.05cm}
%% Target-speaker RNNT 
Target-speaker ASR (TS-ASR)~\cite{denisov2019end,delcroix19_interspeech,moriya2022tsrnnt,zhang2023tsctc,moriya2023tsrnnt_ad,moriya2023tsrnnkd,raj2023tsrnnt_meta} has been proposed as an alternative solution to overlapped ASR. 
TS-ASR recognizes only the target speaker's speech from a mixture using speech enrolled in advance that captures the target speaker's characteristics. 
It naturally avoids output-speaker ambiguity and can limit the decoding to just the target speaker. 
However, for TS-ASR to recognize all speakers in a mixture, the encoder output must be recomputed for each speaker involved using distinct enrolled speech, and each encoder output must be decoded individually. 
Specifically, encoder processing is significantly more computationally expensive than decoder processing~\cite{saon2020alsd,fasoli2022lmfusiontime,yongqiang2023rnnt_joint_ctc_training}. Thus, TS-ASR is not the optimal solution for recognizing all speakers' voices in the mixture simultaneously. 
%Note that this problem also occurs with MT-RNNT using PIT~\cite{SklyarPL21PIT,Sklyar2022mtMSRNNT,sklyar2022sep-rnnt} and HEAT~\cite{lu2021surtorg,Lu2021surt,lu2022epsurt,raj2022mtrnnt,raj2024surt2,raj2024surt_hat}, which require multiple rounds of processing in both the encoder and decoder branches.
Note that this problem also occurs with MT-RNNT using PIT/HEAT, which requires multiple rounds of processing in both the encoder and decoder branches.

%% Speaker attributed MT-RNNT
RNNT-based speaker-attributed ASR has also been proposed as an extension of MT-RNNT-tSOT~\cite{kanda2022SAtSOT}. 
This approach uses an additional speaker encoder/decoder to classify output tokens by speaker. 
Incorporating specific speaker information further improves the performance of multi-talker ASR~\cite{Zhiyun2024sasot_cif}. 
However, it still requires accurate timestamps for the serialization of both target and speaker labels, and the extra encoder/decoder introduces critical delays for streaming ASR. 
Furthermore, for real-world data, speaker information is anonymized and difficult to access.
In this work, we aim to enhance MT-RNNT to recognize multiple speakers while retaining a standard RNNT architecture, without requiring rich alignments, speaker details, or additional encoders.

%%%%%

%\vspace{-0.10cm} 
\section{Baseline systems}
%\vspace{-0.05cm}
Multi-talker ASR recognizes speech from a mixture of $M$ speakers. In this paper, we focus on the two-speaker multi-talker ASR task ($M = 2$), as reported in several studies~\cite{kanda2020sot,lu2021surtorg,Lu2021surt,lu2022epsurt,SklyarPL21PIT,Sklyar2022mtMSRNNT,Kanda2022mtasr,kanda2022SAtSOT,zhang2023tsctc,Zhiyun2024sasot_cif}.
Let $\bm{X}^{\text{mixture}}$ be the input mixture signal of duration $T^{\prime}$; it includes two speakers' voices, denoted as $\bm{X}^{\text{mixture}} = \bm{X}^{\text{spk1}} + \bm{X}^{\text{spk2}}$.
$Y^{\text{spk1}} \in \{1, \dots, K\}^U$ and $Y^{\text{spk2}} \in \{1, \dots, K\}^{U'}$ represent the token sequences associated with each speaker's transcription. $y^{\text{spk}m}_{u} \in \{1, \dots, K\}$ indicates the $u$-th token of the $m$-th speaker in $Y^{\text{spk}m}$.
The vocabulary size, $K$, includes the blank symbol, ``$\phi$''. 

%\vspace{-0.10cm}
\subsection{Single-talker RNNT (Standard RNNT)}
\label{ssec:rnnt}
%\vspace{-0.05cm}
RNNT~\cite{Graves2012} learns the mapping between sequences of different lengths. 
A single-talker speech, $\bm{X}^{\text{spk}1}$, is encoded into $\bm{H}^{\text{enc}} = \left[ \bm{h}^{\text{enc}}_{1}, \dots, \bm{h}^{\text{enc}}_{T} \right]$ of length-$T$ via a feature extractor and encoder network $f^{\text{enc}}(\cdot)$.
$Y^{\text{spk}1}$ is transformed into $\bm{H}^{\text{pred}} = \left[ \bm{h}^{\text{pred}}_{1}, \dots, \bm{h}^{\text{pred}}_{U} \right]$ via prediction network $f^{\text{pred}}(\cdot)$.
These encoded features are then fed to joint network $f^{\text{joint}}(\cdot)$ to obtain the posteriors $\hat{\bm{y}}_{t,u} \in (0,1)^{K}$. 
The above operations are defined as follows:
%\vspace{-0.2cm}
\begin{align}
\bm{h}^{\text{enc}}_{t} &= f^{\text{enc}} (\bm{x}_{t^{\prime}}^{\text{spk}1}; \theta^{\text{enc}}), \\
\bm{h}^{\text{pred}}_{u} &= f^{\text{pred}} (y_{u-1}^{\text{spk}1}; \theta^{\text{pred}}), \\
\hat{\bm{y}}_{t,u}^{\text{spk}1} &= \text{Softmax} \left(f^{\text{joint}} (\bm{h}^{\text{enc}}_{t}, \bm{h}^{\text{pred}}_{u}; \theta^{\text{joint}}) \right),
\label{eq:rnnt}
%\vspace{-0.2cm}
\end{align}
where $\text{Softmax}(\cdot)$ means a softmax operation. 
RNNT outputs three dimensional tensor $\hat{\bm{Y}}^{\text{spk}1} \in (0,1)^{T \times U \times K}$ during training.
The learnable parameters $\theta^{\text{RNNT}} \triangleq [\theta^{\text{enc}}, \theta^{\text{pred}}, \theta^{\text{joint}}]$ are optimized using RNNT loss $\L_{\text{RNNT}}$~\cite{Graves2012}. 
In this study, we retain the original model structure but replace the inputs and outputs with multi-talker variants in the subsequently described MT-RNNT-tSOT and MT-RNNT-AFT. 

\begin{figure}[t]
%\vspace{-0.2cm}
    \centering
    \includegraphics[width=8.2cm]{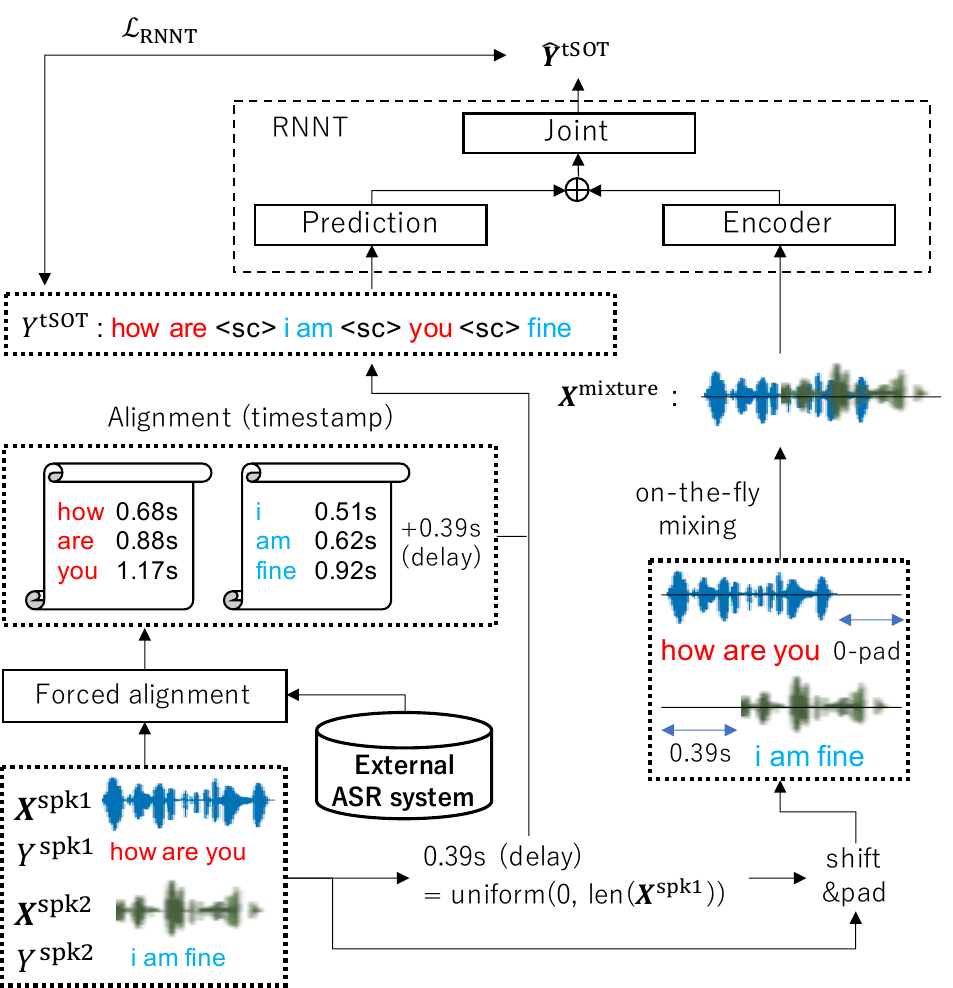}
\vspace{-0.05cm}
 \caption{Training procedure of MT-RNNT-tSOT~\cite{Kanda2022mtasr}. Token-level serialized transcription $Y^{\text{tSOT}}$ is generated from both speakers' transcriptions and their forced alignments obtained from an external ASR system. Both $Y^{\text{tSOT}}$ and its mixture, $\bm{X}^{\text{mixture}}$, are generated on-the-fly with a random delay.}
%\vspace{-0.1cm}
 \label{fig:mt}
\end{figure}

%\vspace{-0.10cm}
\subsection{MT-RNNT with tSOT (MT-RNNT-tSOT)}
\label{ssec:mt}
%\vspace{-0.05cm}
Fig.~\ref{fig:mt} shows the training procedure of the MT-RNNT-tSOT system~\cite{Kanda2022mtasr}. MT-RNNT-tSOT has the same architecture and training procedure as standard RNNT and differs only in the input mixture and its transcriptions. 
The tSOT approach generates training mixture $\bm{X}^{\text{mixture}}$ and labels $Y^{\text{tSOT}}$ on-the-fly~\cite{Kanda2022mtasr} as briefly explained below. 

Two-speaker mixture, $\bm{X}^{\text{mixture}}$, is generated by adding two clean speech signals while ensuring that the second speaker's speech starts after the first speaker. Serialized transcription $Y^{\text{tSOT}}$ is created by sorting all tokens from both speakers based on their timestamps, which are contained in the alignments. 
This process requires accurate timestamps for all tokens, which must be obtained in advance by performing forced alignment on the speech and transcriptions of all speakers using the external ASR system. 
Note that a speaker change token, \texttt{<sc>}, is inserted whenever there is a speaker switch. 

In the training step, since MT-RNNT-tSOT adopts the same architecture as the standard RNNT explained in Section~\ref{ssec:rnnt}, we replace single-talker speech $\bm{X}^{\text{spk}1}$ and its transcription $Y^{\text{spk}1}$ with multi-talker variants, $\bm{X}^{\text{mixture}}$ and $Y^{\text{tSOT}}$, respectively. The joint network of MT-RNNT-tSOT outputs the posteriors probabilities, $\hat{\bm{Y}}^{\text{tSOT}} \in (0, 1)^{T \times (U+U^{\prime}+\alpha) \times (K+1)}$. ``$\alpha$'' represents the number of occurrences of \texttt{<sc>}, and ``$K+1$'' corresponds to the vocabulary size including \texttt{<sc>}.
All parameters, $\theta^{\text{MT-RNNT-tSOT}} \triangleq [\theta^{\text{enc}}, \theta^{\text{pred}}, \theta^{\text{joint}}]$, are optimized with $\mathcal{L}_{\text{RNNT}}$ using $\hat{\bm{Y}}^{\text{tSOT}}$ and $Y^{\text{tSOT}}$. 

For decoding, MT-RNNT-tSOT simultaneously transcribes all speakers' speech in $\bm{X}^{\text{mixture}}$ into a single serial hypothesis $\hat{Y}^{\text{tSOT}}$. 
Although MT-RNNT-tSOT can perform streaming multi-talker ASR, unlike the attentional encoder-decoder (AED) using the utterance-level SOT framework~\cite{kanda2020sot}, it requires accurate alignments from an external pre-trained ASR system. 
Moreover, generating alignments for real mixtures is particularly problematic, resulting in poor alignments, as performing forced alignment is especially challenging. These low-quality alignments lead to the degradation observed in MT-RNNT-tSOT performance. 
Additionally, since the format of the serialized hypothesis $\hat{Y}^{\text{tSOT}}$ is complex, MT-RNNT-tSOT cannot straightforwardly utilize either LM integration~\cite{anjuli2018shallowfusion,meng2021ilme} or the knowledge distillation framework~\cite{knowledge2014hinton} developed for the standard single-talker ASR. 

\begin{figure}[t]
%\vspace{-0.2cm}
 \begin{center}
    %\hspace{-0.4cm}
    \includegraphics[width=8.2cm]{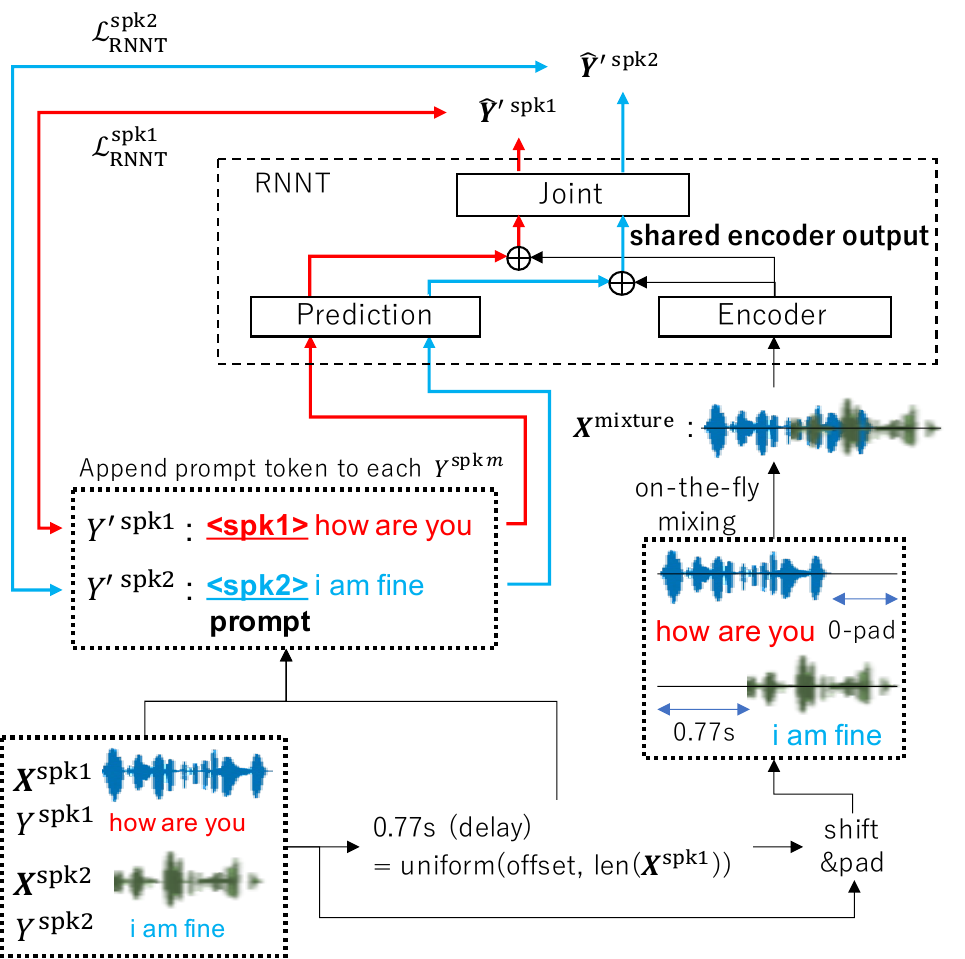}
 \end{center}
\vspace{-0.25cm}
 \caption{Training procedure of MT-RNNT-AFT. 
 MT-RNNT-AFT decodes all speakers' speech in a first-in-first-out manner. Prompt tokens \texttt{<spk}$m$\texttt{>}, which correspond to the sequential order of each speaker’s appearance in mixture $\bm{X}^{\text{mixture}}$, are appended to the beginning of each transcript $Y^{\text{spk}m}$. 
 }
% \vspace{-0.15cm}
 \label{fig:aft}
\end{figure}

%\vspace{-0.025cm}
\section{Proposed methods}
%\vspace{-0.025cm}
\subsection{Alignment-free training for MT-RNNT (MT-RNNT-AFT)}
\label{ssec:aft}
%\vspace{-0.025cm}
In this paper, we propose an alignment-free training scheme for MT-RNNT (MT-RNNT-AFT). This scheme completes the training in a single step and eliminates the need for rich alignments to be generated by an external ASR system. 
Fig.~\ref{fig:aft} shows the procedures for mixture and label generation in MT-RNNT-AFT training. 

MT-RNNT-AFT decodes each speaker's speech in a first-in-first-out manner. 
The mixture generation procedure is the same as that used in MT-RNNT-tSOT, see Section~\ref{ssec:mt}. 
The delay should be set to preserve the order of each speaker’s appearance in the mixture. 
In this paper, the ``offset'' is set to 0.5 seconds, based on the duration of the initial silence within each segment of LibriSpeech~\cite{librispeech}. 

To adhere to the first-in-first-out approach in label generation, we introduce prompt tokens, namely \texttt{<spk1>} for the first speaker, and \texttt{<spk2>} for the second speaker. Each prompt token is appended to the beginning of their respective transcripts, denoted as $Y^{\text{spk}m}$, and the resulting transcript is then named $Y^{\prime \ \text{spk}m}$. 
In the two-speaker case, there are two target labels: $Y^{\prime \ \text{spk1}} \in \{1, \dots, K+2\}^{U+1}$ and $Y^{\prime \ \text{spk2}} \in \{1, \dots, K+2\}^{U^{\prime}+1}$. 
The appearance order information is easier to prepare than obtaining accurate alignments for all training samples, as required by MT-RNNT-tSOT. 
Thus, the AFT scheme can be applied to real data consisting of mixtures and their transcriptions. 

In the training step, MT-RNNT-AFT is trained individually for each speaker. 
The reason is that, in the MT-RNNT-AFT scheme, the number of $Y^{\prime \ \text{spk}m}$ corresponds to $M$ as described above. 
Thus, we feed the mixture $\bm{X}^{\text{mixture}}$ and the transcript $Y^{\prime \ \text{spk}m}$ to the encoder and prediction networks, respectively. 
The joint network of MT-RNNT-AFT computes the predictions $\hat{\bm{Y}}^{\prime \ \text{spk}m}$ for each speaker. Consequently, in the two speaker case ($M=2$), there are two predictions: $\hat{\bm{Y}}^{\prime \ \text{spk1}} \in (0,1)^{T \times (U+1) \times (K+2)}$ and $\hat{\bm{Y}}^{\prime \ \text{spk2}} \in (0,1)^{T \times (U^{\prime}+1) \times (K+2)}$. 
Note that the vocabulary is increased to ``$K+2$'' due to the addition of two prompt tokens.
These predictions are used to calculate the combined loss $\mathcal{L}^{\prime}_{\text{RNNT}} = \sum_{m=1}^{M} \mathcal{L}^{\text{spk}m}_{\text{RNNT}}$, and each loss $\mathcal{L}^{\text{spk}m}_{\text{RNNT}}$ is computed using $\hat{\bm{Y}}^{\text{spk}m}$ and $Y^{\text{spk}m}$.  
All parameters, $\theta^{\text{MT-RNNT-AFT}} \triangleq [\theta^{\text{enc}}, \theta^{\text{pred}}, \theta^{\text{joint}}]$, are optimized with $\mathcal{L}_{\text{RNNT}}^{\prime}$.

In decoding, MT-RNNT-AFT recognizes all speakers' voices in a first-in-first-out manner by processing the mixture through the encoder just once. 
By inputting the corresponding prompt token at the beginning, the decoder, consisting of the prediction and joint networks, outputs each speaker's hypothesis from the shared encoder output. 
%Beam search can be performed in parallel by batching decoder processing~\cite{saon2020alsd} for all speakers, thanks to the use of identical parameters. 
Beam search can be performed in parallel by batching decoder processing~\cite{saon2020alsd} for all speakers. 
%If multi-threading is available, the beam size can remain unchanged; otherwise, it can be reduced to equalize computational costs.
Thus, the processing of the encoder and beam search, including the decoder, is completed in just one pass for all speakers, thanks to the fully shared parameters and the use of the shared encoder output.
Therefore, the total computational cost of MT-RNNT-AFT is much lower than that of TS-ASR and MT-RNNT using PIT/HEAT. 
This is because encoder processing is significantly more computationally expensive than decoder processing~\cite{saon2020alsd,fasoli2022lmfusiontime,yongqiang2023rnnt_joint_ctc_training}. 
Moreover, TS-ASR and MT-RNNT using PIT/HEAT require multiple invocations of both the encoder and decoder modules.\footnote{We also applied PIT/HEAT to the standard RNNT architecture without additional encoders or decoders for MT-RNNT training, but the training loss failed to converge as the identical parameters lacked speaker identifiers. The proposed prompt tokens, which identify each speaker, address this issue.}

%%%%

%\vspace{-0.25cm}
\subsection{Self-knowledge distillation for MT-RNNT-AFT}
\label{sssec:kd}
%\vspace{-0.15cm}

In this paper, we also propose a self-knowledge distillation (KD) approach to further enhance MT-RNNT-AFT.
Multiple single-talker speech $\bm{X}^{\text{spk}m}$ is naturally available for MT-RNNT-AFT training due to the simulated on-the-fly mixture generation process.
We exploit the parallel speech data, i.e., $\bm{X}^{\text{spk}m}$ and $\bm{X}^{\text{mixture}}$, in our KD framework, similar to~\cite{zhang19i_interspeech,moriya2023tsrnnkd}.  
The training process consists of three steps. 
First, we obtain pseudo labels $\hat{\bm{Y}}^{\text{spk}m}$ of the $m$-th speaker by processing each single-talker ASR data, $\bm{X}^{\text{spk}m}$ and $Y^{\prime \ \text{spk}m}$, with MT-RNNT-AFT before mixing. 
Then, we obtain predictions $\hat{\bm{Y}}^{\prime \ \text{spk}m}$ by processing multi-talker ASR data, $\bm{X}^{\text{mixture}}$ and $Y^{\prime \ \text{spk}m}$, with MT-RNNT-AFT. 
%Finally, we compute the KD loss $\mathcal{L}^{\text{spk}m}_{\text{KD}}$ between $\hat{\bm{Y}}^{\text{spk}m}$ and $\hat{\bm{Y}}^{\prime \ \text{spk}m}$. 
%The KD loss, $\mathcal{L}_{\text{KD}}$, and the combined loss, $\mathcal{L}_{\text{RNNT+KD}}$, with $\mathcal{L}^{\prime}_{\text{RNNT}}$ for MT-RNNT-AFT training are defined as follows:
Finally, we compute each speaker's KD loss, $\mathcal{L}^{\text{spk}m}_{\text{KD}}$, using $\hat{\bm{Y}}^{\text{spk}m}$ and $\hat{\bm{Y}}^{\prime \ \text{spk}m}$, and then sum them into $\mathcal{L}_{\text{KD}}$. 
The total KD loss, $\mathcal{L}_{\text{KD}}$, and the combined loss, $\mathcal{L}_{\text{RNNT+KD}}$, with $\mathcal{L}^{\prime}_{\text{RNNT}}$ for MT-RNNT-AFT training are defined as follows:
\begin{align}
  \mathcal{L}_{\text{KD}} &=  - \sum_{m=1}^{M} \sum_{t=1}^{T} \sum_{u=1}^{U} \sum_{k=1}^{K+M} \hat{y}^{\text{spk}m}_{t,u,k} \ \log \ \hat{y}^{\prime \ \text{spk}m}_{t,u,k}, \\
  \mathcal{L}_{\text{RNNT+KD}} &= \mathcal{L}^{\prime}_{\text{RNNT}} + \lambda \mathcal{L}_{\text{KD}},
  \label{eq:KD}
%\vspace{-0.1cm}
\end{align}
where $\hat{y}^{\text{spk}m}_{t,u,k}$ and $\hat{y}^{\prime \ \text{spk}m}_{t,u,k}$ correspond to the $k$-th class probability of $\hat{\bm{Y}}^{\text{spk}m}$ and $\hat{\bm{Y}}^{\prime \ \text{spk}m}$ at the $t$-th time and $u$-th label steps of $m$-th speaker, respectively. 
$\lambda$ is the weight of $\L_{\text{KD}}$. 
We expect that the frame-level pseudo labels from MT-RNNT-AFT, generated using single-talker ASR data, will improve the model's training stability and guide alignment when processing multi-talker ASR data.

%%%%

%%%%
\begin{table*}[t]
\centering
\caption{
cpWERs [\%] of each ASR system on LibriSpeechMix. 
``Additional info.'' columns denote the additional information required for training/decoding. 
``Align'': forced alignment generated by an external pre-trained ASR system.
``Spk'': speaker information, e.g., speaker IDs and enrollment details. 
``Enc'': an additional encoder that operates separately from the main ASR encoder.}
\setlength{\tabcolsep}{2pt}
\vspace{-0.25cm}
\label{tab:comp}
\subfloat[Offline results]{%
\begin{minipage}{0.49\linewidth}
\vspace{-0.15cm}
\resizebox{\linewidth}{!}{%
\begin{tabular}{@{}lcccccccc@{}}
\bhline{1pt}
\multirow{2}{*}{System} & \multirow{2}{*}{\begin{tabular}[c]{@{}c@{}}\# of\\ param.\end{tabular}} & \multirow{2}{*}{\begin{tabular}[c]{@{}c@{}}Latency\\ {[}ms{]}\end{tabular}} & \multicolumn{3}{c}{Additional info.}                   & \multirow{2}{*}{LM} & \multicolumn{2}{c@{}}{cpWER} \\ \cline{4-6} \cline{8-9}  
                        &                                   &                                                                             & Align & Spk & Enc &  & 1spk            & 2spk           \\ \hline
%MT-AED-SOT~\cite{kanda2021SASOT}       & Offline                     & 129M                                                                    & $\infty$                                                                           & -         & -          & -     & -           & 3.6             & 4.9            \\
MT-AED-SASOT~\cite{kanda2021SASOT}    & 142M                                                                    & $\infty$                                                                           & -        &\checkmark  &\checkmark  & -      & 3.3             & 4.3            \\
TS-CTC~\cite{zhang2023tsctc}     & 88M                                                                     & $\infty$                                                                           & -         &\checkmark  &\checkmark   & -     & -               & 4.2            \\
MT-RNNT-tSOT~\cite{Kanda2022mtasr}            & 139M                                                                    & 2560                                                                        &\checkmark  & -          & -         & -       & 3.3             & 4.4            \\
MT-CIF-tSOT~\cite{Zhiyun2024sasot_cif}         & 136M                                                                    & $\infty$                                                                           &\checkmark &\checkmark  &\checkmark  & -      & 2.5             & 3.4            \\ \hdashline
Single-talker RNNT (ours)    & 120M                                                                    & $\infty$                                                                           & -        & -          & -        & -        & 2.7             & 64.5            \\
MT-RNNT-tSOT (ours)    & 120M                                                                    & $\infty$                                                                           &\checkmark & -          & -          & -      & 2.6             & 4.0            \\
MT-RNNT-AFT (proposed)      & 120M                                                                    & $\infty$                                                                           & -         & -          & -       & -         & 2.8             & 3.9            \\
+KD (proposed)          & 120M                                                                    & $\infty$                                                                           & -         & -          & -        & -        & 2.6             & 3.7            \\ %\hline 
\ \ +ILME (proposed)          & +36M                                                                    & $\infty$                                                                           & -         & -          & -      & \checkmark   & \textbf{2.4}             & \textbf{3.4}            \\ \bhline{1pt}
\end{tabular}%
}%
\end{minipage}%
}
\hfill
\subfloat[Streaming results ($\leq 1000$ms latency)]{%
\begin{minipage}{0.49\linewidth}
\vspace{-0.15cm}
\resizebox{\linewidth}{!}{%
\begin{tabular}{@{}lcccccccc@{}}
\bhline{1pt}
\multirow{2}{*}{System} & \multirow{2}{*}{\begin{tabular}[c]{@{}c@{}}\# of\\ param.\end{tabular}} & \multirow{2}{*}{\begin{tabular}[c]{@{}c@{}}Latency\\ {[}ms{]}\end{tabular}} & \multicolumn{3}{c}{Additional info.}                   & \multirow{2}{*}{LM} & \multicolumn{2}{c@{}}{cpWER} \\ \cline{4-6} \cline{8-9}
                        &                                                                         &                                                                             & Align & Spk & Enc &  & 1spk            & 2spk           \\ \hline
MT-RNNT-SURT~\cite{lu2022epsurt}            & 85M                                                                     & 1000                                                                        & -         & -          &\checkmark    & -    & -               & 9.1            \\
%MT-RNNT-PIT~\cite{SklyarPL21PIT}             & Streaming                         & 81M                                                                     & 30                                                                          & -         & -          &\checkmark   & -     & 7.6             & 10.2           \\
MT-RNNT-PIT~\cite{Sklyar2022mtMSRNNT}             & 81M                                                                     & 30                                                                          & -         & -          &\checkmark        & -        & -      & 8.8            \\
%MT-RNNT-tSOT~\cite{Kanda2022mtasr}            & Streaming                         & 82M                                                                     & 160                                                                         &\checkmark & -          & -        & -        & 4.9             & 6.9            \\
MT-RNNT-tSOT~\cite{Kanda2022mtasr}            & 82M                                                                     & 640                                                                         &\checkmark & -          & -        & -        & 4.2             & 6.2            \\
MT-RNNT-tSOT~\cite{Kanda2022mtasr}            & 139M                                                                    & 160                                                                         &\checkmark & -          & -         & -       & 4.3             & 6.2            \\ \hdashline
Single-talker RNNT (ours)     & 120M                                                                    & 640                                                                         & -           & -          & -         & -       & 4.1             & 66.6            \\
MT-RNNT-tSOT (ours)     & 120M                                                                    & 640                                                                         &\checkmark & -          & -         & -       & 4.2             & 6.5            \\
MT-RNNT-AFT (proposed)      & 120M                                                                    & 640                                                                         & -         & -          & -       & -         & 4.9             & 7.4            \\
+KD (proposed)              & 120M                                                                    & 640                                                                         & -         & -          & -        & -        & 4.1             & 6.7            \\ %\bhline{1pt}
\ \ +ILME (proposed)              & +36M                                                                    & 640                                                                         & -         & -          & -      & \checkmark          & \textbf{4.0}             & \textbf{6.3}            \\ \bhline{1pt}
\end{tabular}%
}%
\end{minipage}%
}
\vspace{-0.05cm}
\end{table*}

%\vspace{-0.15cm}
\section{Experimental evaluations}
\label{sec:result}
%\vspace{-0.05cm}
\subsection{Data}
\label{ssec:data}
%\vspace{-0.15cm}

We used the LibriSpeech corpus~\cite{librispeech} for training, and LibriSpeechMix~\cite{kanda2020sot}\footnote{\url{https://github.com/NaoyukiKanda/LibriSpeechMix}} for the development and evaluation sets. 
We used simulated mixtures generated on-the-fly, as described in Section~\ref{ssec:mt}. 
Volume and speed perturbation~\cite{povey2015sp} and SpecAugment~\cite{specaugment} were applied to the speech after on-the-fly mixing during training. 
The proportions of single-talker and two-speaker ASR data during training were 50\% each. 
For tSOT label creation, the forced alignments were generated by using the Montreal Forced Aligner~\cite{mcauliffe2017MFA}. 
We adopted the 1k subwords determined by SentencePiece~\cite{kudo2018subword}. 
We performed experiments using the ESPnet~\cite{espnet}. 
We measured model performance using the concatenated minimum-permutation word error rate (cpWER)~\cite{chime6} for both single-talker (1spk) and two-speaker (2spk) ASR tasks.

%\vspace{-0.10cm}
\subsection{System configuration}
\label{ssec:asrsystem}
%\vspace{-0.05cm}

We used an 80-dimensional log Mel-filterbank, extracted every 10ms, as the input feature of ASR models. 
We adopted Conformer (L)~\cite{anmol2020conformer}, where batch normalization was replaced by layer normalization; kernel size was reduced from 31 to 15.
The encoder contained a two-layer 2D convolutional neural network (CNN) followed by 17 Conformer blocks.  
The prediction network had a 640-dimensional long short-term memory (LSTM) layer. 
The joint network consisted of a 512-dimensional feed-forward network. 

For the streaming experiments, we constructed a variant of the offline system configuration, with only the offline encoder replaced by a chunkwise Conformer encoder~\cite{chen2021lcconformer}. 
Both the current and history chunk sizes of the streaming Conformer encoder were set to 60 frames, so the algorithmic latency was $640\text{ms} = 600\text{ms} + 40\text{ms}$, with 40ms added due to the number of CNN lookahead frames. 
While the parameters of the offline Conformer model were randomly initialized, the streaming Conformer parameters were initialized with those from the trained offline Conformer.

For the MT-RNNT-AFT, we used on-the-fly internal LM estimation (ILME) during decoding~\cite{meng2021ilme}. The LM consisted of a four-layer LSTM with 1024 cells, and was trained using a large amount of text data following the LibriSpeech recipe. The ILM was jointly trained with MT-RNNT-AFT as detailed in~\cite{meng2021ilmt,moriya2022rnntadlm,moriya2023rnntss}. 
%The LM weights were tuned on the development set.

For the training process, we utilized the AdamW optimizer along with a warmup learning rate scheduler; a peak learning rate of 1.5e-3 was reached after 25k warmup steps, and all models were run for a total of 200 epochs each. 
For the MT-RNNT-AFT model, we set $\lambda$ to $0.001$ when we applied KD loss $\mathcal{L}_{\text{KD}}$, described in Section~\ref{sssec:kd}, with its application starting at the 180th epoch.
%This is because, at the beginning of MT-RNNT-AFT training, the output used as pseudo labels may not be very accurate.
%We also built a single-talker RNNT using only the plain LibriSpeech corpus. 
The minibatch size was set to 256 in all experiments. 
For decoding, we utilized alignment-length synchronous decoding~\cite{saon2020alsd} with a beam size of 16.

%% Experimental results

\subsection{Offline results} 
%\vspace{-0.05cm}
Table~\ref{tab:comp} (a) shows the offline results. 
The check marks in Table~\ref{tab:comp} (a) indicate the additional information utilized for training and/or decoding. 
The results from the literature are displayed above the dashed line. 
The values written below the dashed line present our reproduced MT-RNNT-tSOT and our proposal, MT-RNNT-AFT. 

First, the single-talker RNNT model struggled to recognize speech in a mixture. 
Our reproduced MT-RNNT-tSOT achieved better cpWERs than the original MT-RNNT-tSOT~\cite{Kanda2022mtasr}. 
Thus, our reproduced MT-RNNT-tSOT establishes a state-of-the-art baseline; it utilizes the standard RNNT architecture, without any speaker information or additional encoder.
MT-RNNT-AFT achieved performance comparable to that of MT-RNNT-tSOT, despite not using any additional information. 
By applying the KD loss proposal during training, the performance of MT-RNNT-AFT was further enhanced, allowing it to outperform MT-RNNT-tSOT. 
%Therefore, MT-RNNT-AFT achieved the best performance while retaining the same architecture as standard RNNT. 
Therefore, MT-RNNT-AFT achieved the best performance while retaining the standard RNNT architecture. 

Additionally, while LM integration is challenging for MT-RNNT-tSOT as its complex hypotheses contain mixed words from all speakers, MT-RNNT-AFT can be naturally integrated with an external LM. This is because each hypothesis individually contains the words spoken by each speaker. We applied ILME to MT-RNNT-AFT trained with KD, and it achieved performance comparable to the state-of-the-art as reported in~\cite{Zhiyun2024sasot_cif}, which requires additional information such as rich alignments, specific speaker information, and an extra encoder. 
 
%\vspace{-0.1cm}
\subsection{Streaming results} 
%\vspace{-0.05cm}
Next, we performed streaming experiments; the results are shown in Table~\ref{tab:comp} (b). 
We observed that MT-RNNT-AFT operates effectively in streaming mode. However, its performance failed to match that of MT-RNNT-tSOT. 
%The degradations were caused by insertion errors that occurred during longer durations of inactive speech, including silence or speech from other speakers. 
%The reason is that streaming MT-RNNT-AFT lacks a mechanism to carry speaker information across chunks or access larger look-ahead frames, such as tracking the order of each speaker's appearance in the mixture. 
The deficiencies were caused by deletion errors in the 1spk task and insertion errors in the 2spk task. These errors occurred during a longer duration of inactive speech, including silence or speech from other speakers. 
%The reason is that streaming MT-RNNT-AFT lacks a mechanism to carry speaker information across chunks or access larger look-ahead frames, such as tracking the order of each speaker's appearance in the input speech. 
The reason is that streaming MT-RNNT-AFT lacks a mechanism to carry speaker information across chunks or access larger look-ahead frames, such as tracking the order of each speaker's appearance and their presence in the next input chunk.
Notably, KD using frame-level pseudo labels, which include not only posteriors but also speaker activity information, improved the results of streaming MA-RNNT-AFT. The results were comparable to those of MT-RNNT-tSOT, which utilizes rich alignment. 

We also applied ILME to MT-RNNT-AFT trained with KD and found its performance to match that of state-of-the-art alternatives, as reported in~\cite{Kanda2022mtasr}.
Despite severe challenges by performing the task without rich alignment, speaker information, or an additional encoder, MT-RNNT-AFT achieved performance comparable to that of MT-RNNT-tSOT in both offline and streaming modes.

%%%%
\begin{table}[t]
\centering
\caption{
Comparison of cpWERs [\%] for each MT-RNNT on LibriSpeechMix (1spk/2spk) across different beam sizes.
}
\vspace{-0.15cm}
\label{tab:beam}
\setlength{\tabcolsep}{0.15cm}
\begin{tabular}{@{}llccccc@{}}
\bhline{1pt}
\multirow{2}{*}{System}                                                    & \multirow{2}{*}{Mode} & \multicolumn{5}{c@{}}{Beam size} \\
                                                                           &                       & 1          & 2         & 4         & 8         & 16        \\ \hline
\multirow{2}{*}{MT-RNNT-tSOT}                                              & Offline               & 2.7/5.7    & 2.7/4.1   & 2.6/4.1   & 2.6/4.0   & 2.6/4.0   \\
                                                                           & Streaming             & 4.4/9.2    & 4.2/6.7   & 4.2/6.6   & 4.2/6.5   & 4.2/6.5   \\ \hline
\multirow{2}{*}{\begin{tabular}[c]{@{}l@{}}MT-RNNT-AFT\\ +KD (proposed)\end{tabular}} & Offline               & 2.7/4.1    & 2.6/3.7   & 2.6/3.7   & 2.6/3.7   & 2.6/3.7   \\
                                                                           & Streaming             & 4.3/7.2    & 4.2/6.7   & 4.2/6.7   & 4.1/6.7   & 4.1/6.7   \\ \bhline{1pt}
\end{tabular}
\vspace{-0.05cm}
\end{table}
%%%%

%\vspace{-0.05cm}
\subsection{Effect of beam size in inference of each MT-RNNT}
%\vspace{-0.025cm}

Although the above experiments consistently used a beam size of 16, the multi-threaded decoder processing may not be available, even though batching is supported. 
In that case, MT-RNNT-AFT reduces the beam size to equalize computational costs.
Thus, we investigated various beam sizes and their effects on the cpWERs of each MT-RNNT model; the results are detailed in Table~\ref{tab:beam}. 
%From Table~\ref{tab:beam}, when the beam size for MT-RNNT-tSOT was set to 4 and for MT-RNNT-AFT to 2, MT-RNNT-AFT matched the performance of MT-RNNT-tSOT, even with single-threaded decoding, and without any noticeable degradation at the smallest beam size.
From Table~\ref{tab:beam}, when the beam size for MT-RNNT-tSOT was set to 4 and for MT-RNNT-AFT to 2, MT-RNNT-AFT matched the performance of MT-RNNT-tSOT without significant degradation at the smallest beam size.

%\vspace{-0.0125cm}
\section{Conclusion}
\label{ssec:conclusions}
%\vspace{-0.0125cm}
%\vspace{0.05cm}
We have proposed MT-RNNT-AFT, an alignment-free training enhanced MT-RNNT that can be trained without requiring rich alignments while retaining the standard RNNT architecture. 
We introduced a prompt token that informs the MT-RNNT-AFT which speaker to recognize in the mixture. 
This procedure simplifies the decoding process, resulting in a much simpler training approach, while also enabling the use of KD and ILME. 
MT-RNNT-AFT achieved performance comparable to that of MT-RNNT-tSOT, which requires rich alignments. 
Moreover, offline MT-RNNT-AFT matched the performance of the state-of-the-art alternatives, while the latter requires rich alignments, speaker details, and an additional encoder.

\newpage

\clearpage
\bibliographystyle{IEEEbib-abbrev}
\bibliography{mybib,refs}

\end{document}